\documentstyle[12pt]{article}
\def\slash#1{#1\hskip -0.5em/}
\def\bbar{\overline{B}}
\title{QCD Corrected $1/m_b$ Contributions to $B\bbar$--Mixing\thanks{%
Darmstadt Preprint IKDA 92/36, paper hep-ph/9211333}}
\author{{\sc Wolfgang Kilian \quad
             Thomas Mannel }
        \vspace*{5mm} \\
        Institut f\"ur Kernphysik, \\
        Technische Hochschule Darmstadt, \\
        Schlossgartenstr. 9, D--6100 Darmstadt  \\
        Germany
	\vspace*{5mm}}
\begin{document}
\maketitle
\vfill
\begin{abstract}
\noindent
We calculate the QCD corrected effective Hamiltonian
for $B\bbar$--Mixing
in heavy quark effective theory including
corrections of the order $\Lambda_{QCD} / m_b$.
The matrix elements of the subleading operators
are estimated using the vacuum insertion assumption.
We show that the major part of the subleading corrections may
be absorbed into the heavy meson decay constant $f_B$; the remaining
corrections are only due to QCD effects and give an enhancement
of $\Delta M$ of 5\%.
\end{abstract}
\thispagestyle{empty}
\newpage
\setcounter{page}{1}
\section{Introduction}
Oscillations between particle and antiparticle have been observed
first in the neutral Kaon system \cite{La56} and later also in the
neutral $B$ meson system \cite{Al87,Ar89}. Such oscillations are
predicted by the standard model and proceed through the so called
box diagrams depicted in fig. 1.
Since the top quark as well as the $W$ boson are heavy compared to the
$B$ meson it is convenient to describe the oscillations in
an effective theory, where the top quark and the $W$ boson are
integrated out.

We shall concentrate in this paper on the oscillations in the
neutral $B$ meson system. The effective Hamiltonian, which is obtained
after integrating out the top quark and the $W$ boson,
consists of only one operator
\begin{equation} \label{heff}
H_{\rm eff} = \frac{G_F^2}{\pi^2} |V_{tb}^* V_{td}|^2 m_t^2
          \Phi\left(\frac{m_t^2}{M_W^2}\right)
          {\cal O}_0
\end{equation}
where
\begin{equation}\label{O_0}
{\cal O}_0 = (\bar{d}_L\gamma_\mu b) \,(\bar{d}_L\gamma^\mu b)
\end{equation}
($d_L$ denotes the left-handed component of the $d$ quark field)
and $\Phi$ is a function which arises from integrating out the
top quark and the $W$ boson at the same scale \cite{IL81}
\begin{equation}
\Phi(x) = \frac{1}{4} + \frac{9}{4(1-x)} - \frac{3}{2(1-x)^2}
       - \frac{3}{2}\frac{x^2 \ln x}{(1-x)^3}
\end{equation}

To evaluate the mass shift $\Delta M$ relevant for the oscillations,
one needs to calculate
the matrix element of this effective interaction between a neutral
$B$ meson and its antiparticle; it is given by
\begin{eqnarray}\label{delta-m-1}
\Delta M &=&
\frac{1}{2m_B}\langle B^0 | H_{\rm eff} | \bbar^0 \rangle \nonumber \\
        &=& \frac{G_F^2}{2\pi^2} |V_{tb}^* V_{td}|^2 m_t^2
            \Phi\left(\frac{m_t^2}{M_W^2}\right)
	    \frac1{m_B}
	    \langle B^0 | {\cal O}_0 | \bbar^0 \rangle \\
        &=& \frac{G_F^2}{6 \pi^2} |V_{tb}^* V_{td}|^2 m_t^2
            \Phi\left(\frac{m_t^2}{M_W^2}\right) \eta_{\rm QCD} (\mu)\,
            f^2_B B_B (\mu)\,m_B\nonumber
\end{eqnarray}
where $f_B$ is the $B$ meson decay constant defined by
\begin{equation}
\langle 0 | \bar{d} \gamma_\mu \gamma_5 b | \bbar^0(p) \rangle = if_B p_\mu.
\end{equation}
The coefficient $\eta_{\rm QCD}$ contains the short distance
contribution of the QCD corrections which is calculable in perturbation theory.
It depends on the factorization
scale $\mu$ at which the short distance contributions become separated
from the long distance ones; the non-perturbative long distance contributions
are contained in the ``bag parameter'' $B_B$, which depends in such a
way on $\mu$ that the matrix element of the Hamiltonian is scale
independent.

As an estimate for the order of magnitude of the effect the so called
``vacuum insertion'' assumption has been frequently used in the
past, which corresponds to the case $B_B = 1$. However, since
$B_B$ is scale dependent this assumption has to be supplemented
by a statement about the scale where it is assumed to be valid.

The short distance contribution from scaling down to a scale
$m_b \le \mu \le M_W$ has been calculated in leading
logarithmic approximation some time ago
\cite{Ha79} and yields
\begin{equation}\label{full-scaling}
\eta_{\rm QCD} (\mu) = \left( \frac{\alpha_s (\mu)}{\alpha_s(M_W)}
                   \right)^{-6/23}\quad\quad m_b \leq \mu \leq M_W,
\end{equation}
which gives a correction factor of $\eta_{\rm QCD} \sim 0.85$ at the
scale of the $b$ quark. The corrections induced by subleading logarithms
wich are to be considered in conjunction with the $O(\alpha_s)$
matching conditions at the $W$ scale
have also been calculated \cite{BJW90}.

If one could get some information on the parameter $B_B$ at a scale
of the order of the $B$ quark mass, one could in fact predict the
amplitude of the oscillations
as a function of the CKM matrix elements and the top quark mass.
However, at the present time the
only non-perturbative information available is from lattice methods
\cite{Lattice}, which gives some information on the parameter
$f_B^2 B_B$ at low scales\footnote{%
   In fact,
   the lattice results indicate that the ``bag parameter'' $B_B$
   at these low scales
   is indeed close to the value $B_B = 1$ used in the vacuum insertion
   assumption \cite{Lattice}.}.
Thus it is
desirable to scale further down to these very low scales. This is
possible by switching at the point $\mu = m_b$ to heavy quark
effective theory \cite{HQET}, in which the $b$ quark is integrated out
and replaced by a static color source. The leading logarithmic
result for the scaling $\mu \le m_b$, i.~e.\ the resummation of
logarithms $(\alpha_s\ln(m_b/\mu))^n$, has been calculated in
\cite{VS89}. The result is
\begin{eqnarray}
  \eta_{\rm QCD}(\mu) &=& \left( \frac{\alpha_s(\mu)}{\alpha_s(m_b)} \right)
		^{12/25} \eta_{\rm QCD}(m_b)
		\quad\quad m_c \leq \mu \leq m_b \\
  \eta_{\rm QCD}(\mu) &=& \left( \frac{\alpha_s(\mu)}{\alpha_s(m_c)} \right)
		^{12/27} \eta_{\rm QCD}(m_c)
		\quad\quad \mu \leq m_c.
\end{eqnarray}
This result is indeed very peculiar, since the anomalous dimension
of the four fermion operator ${\cal O}_0$ is the sum of the anomalous
dimensions of the two left handed currents involved. In fact,
if factorization like e.~g. vacuum insertion assumption
takes place, one would expect this behaviour of
the anomalous dimension to all orders.
However,
the QCD scaling below $m_b$ has been calculated even to subleading
order \cite{Gim92} and it seems that this factorization does not hold for the
subleading result.

Heavy quark effective theory is a systematic expansion in inverse
powers of the heavy quark mass. The purpose of the present paper is
to extend the existing calculations of the scaling below $m_b$ by
including terms of the order $1/m_b$ into the effective Hamiltonian.
The motivation is twofold. Firstly, there are indications from
the lattice that the $1/m_b$ corrections could be large, even for the
$b$ quark. Secondly, this is the first attempt to deal with the
$1/m_b$ corrections in a non-leptonic process, of which the
Hamiltonian for $B \bbar$ mixing is certainly the easiest to
deal with.

In section 2 we shall discuss an appropriate operator basis for the
$1/m_b$ corrections to $B\bbar$ mixing and calculate the anomalous
dimension matrix and the Wilson coefficients at some low scale
$\mu$. In section 3 we shall estimate the $1/m_b$ effects by
using the vacuum insertion assumption and the results on the
$1/m$ corrections for the heavy meson decay constants obtained
from QCD sum rules \cite{Ne91,Ba91}.

\section{The Operator Basis in the Order $1/m_b$}

To leading and subleading order in $1/m$, the effective Lagrangian for
a heavy quark moving with velocity $v$ is given by
\cite{MRR91} (we use $D_\mu = \partial_\mu - i g A_\mu^a T^a$)
\begin{eqnarray}\label{HQET-Lagrangian}
  {\cal L} &=&
  \bar h_v^+ i v\cdot D h_v^+ + \bar\rho_v^+ h_v^+ +
  \bar h_v^+ \rho_v^+\\
  && \nonumber
  + \frac{1}{2m}
  \left(\bar h_v^+ i\slash D^\perp + \bar R^+_v\right)
  \left(i\slash D^\perp h^+_v + R_v^+\right)
\end{eqnarray}
where the transverse derivative $D - \slash v v\cdot D$
is denoted as $D^\perp$,
the heavy quark field is defined by
\begin{equation}
  h_v^+(x) = e^{i m v\cdot x} \frac{1 + \slash v}{2} b(x),
\end{equation}
and the sources $\rho_v^+$ and $R_v^+$ for the heavy quark and antiquark
fields with velocity $v$ have
been retained. Since there is also a heavy antiquark involved, one has to
use also the effective Lagrangian for an antiquark which can be obtained
from (\ref{HQET-Lagrangian}) just by replacing $v$ with $-v$ and $h_v^+$ with
\begin{equation}
  h_v^-(x) = e^{-i m v\cdot x} \frac{1 - \slash v}{2} b(x),
\end{equation}
along with the proper replacements of the sources:
$\rho_v^+ \to \rho_v^-$, $R_v^+ \to R_v^-$.

Using the quark-type parameterization for one factor in the effective
Hamiltonian (\ref{O_0})
and the antiquark-type parameterization in the other factor
(using the nomenclature of \cite{MRR91})
one obtains the effective Hamiltonian density at the matching scale
$\mu = m_b$ in terms of heavy quark fields
\begin{eqnarray}\label{HQET-heff}
  && H_{\rm eff}
  = \frac{G_F^2}{\pi^2} |V_{tb}^* V_{td}|^2 m_t^2
          \Phi\left(\frac{m_t^2}{M_W^2}\right)
  \\
  && \times \left[
  \eta_{\rm QCD}(m_b)\, {\cal O}'_0(m_b)
   + \frac{1}{m_b}\left\{ a_1(m_b)\, X_1^+(m_b)
   + \sum_{i=1}^3 c_i(m_b)\,{\cal O}_i^+(m_b)  \right\}\right]. \nonumber
\end{eqnarray}
At the matching scale there are two local operators
\begin{eqnarray}\label{match}
  {\cal O}'_0 &=& (\bar d_L \gamma_\mu h_v^+)\,(\bar d_L \gamma^\mu h_v^-)
  \\  \nonumber
  X_1^+      &=& (\bar d_L \gamma_\mu i \slash D^\perp h_v^+)\,
	(\bar d_L \gamma^\mu h_v^-) + (\bar d_L \gamma_\mu h_v^-)\,
	(\bar d_L \gamma^\mu i \slash D^\perp h_v^-),
\end{eqnarray}
and three nonlocal operators ${\cal O}_i^+$,
which are time-ordered products originating
from the $1/m$ piece of the Lagrangian:
\begin{eqnarray}
  {\cal O}_1^+ &=& i\int d^4 x\,
	{\rm T}\left[ {\cal O}'_0(0)\, ( \bar h_v^+ ( iv\cdot D)^2 h_v^+
		+ \bar h_v^- (i v\cdot D)^2 h_v^- )(x) \right] \nonumber\\
  {\cal O}_2^+ &=& i\int d^4 x\,
	{\rm T}\left[ {\cal O}'_0(0)\, ( \bar h_v^+ ( iD)^2 h_v^+
		+ \bar h_v^- (iD)^2 h_v^- )(x) \right] \\
  {\cal O}_3^+ &=& i\int d^4 x\,
	{\rm T}\left[ {\cal O}'_0(0)\, ( \bar h_v^+
			\frac{g}{2} \sigma^{\mu\nu}G_{\mu\nu} h_v^+
		+ \bar h_v^- \frac{g}{2} \sigma^{\mu\nu} G_{\mu\nu} h_v^- )(x)
	\right].\nonumber
\end{eqnarray}
The coefficients of the $1/m_b$ terms are
\begin{eqnarray}\label{initial}
  &a_1(m_b) = \frac12\eta_{\rm QCD}(m_b),
  \\ \nonumber
  &c_2(m_b) = c_3(m_b) = -c_1(m_b) = \frac12\eta_{\rm QCD}(m_b),
\end{eqnarray}

In order to calculate the leading logarithmic QCD corrections to this effective
Hamiltonian, one first has to set up a basis of operators bearing the
correct dimension and quantum numbers. Since there are three independent
momenta in the transition, and five independent gamma matrix structures
possible, one arrives at a basis of 15 local operators:
\begin{eqnarray}
  P_1 &=& (\bar d_L iv\cdot D h_v^+)\,(\bar d_L h_v^-) \nonumber\\
  P_2 &=& (\bar d_L i\slash D h_v^+)\,(\bar d_L h_v^-) \nonumber\\
  P_3 &=& (\bar d_L iD_\mu h_v^+)\,(\bar d_L \gamma^\mu h_v^-) \nonumber\\
  P_4 &=& (\bar d_L \gamma_\mu iv\cdot D h_v^+)\,(\bar d_L \gamma^\mu h_v^-)
\nonumber\\
  P_5 &=& i \epsilon_{\lambda\mu\nu\rho} v^\lambda
	(\bar d_L iD^\mu\gamma^\nu h_v^+)\,(\bar d_L \gamma^\rho h_v^-) \nonumber\\
  \nonumber\\
  Q_1 &=& (\bar d_L h_v^+)\,(\bar d_L iv\cdot D h_v^-) \nonumber\\
  Q_2 &=& (\bar d_L \gamma_\mu h_v^+)\,(\bar d_L iD^\mu h_v^-) \nonumber\\
  Q_3 &=& (\bar d_L h_v^+)\,(\bar d_L i\slash D h_v^-) \\
  Q_4 &=& (\bar d_L \gamma_\mu h_v^+)\,(\bar d_L \gamma^\mu iv\cdot D h_v^-)
\nonumber\\
  Q_5 &=& i \epsilon_{\lambda\mu\nu\rho} v^\lambda
	(\bar d_L \gamma^\nu h_v^+)\,(\bar d_L iD^\mu\gamma^\rho h_v^-) \nonumber\\
  \nonumber\\
  R_1 &=& \left[iv\cdot \partial (\bar d_L h_v^+)\right](\bar d_L h_v^-)
\nonumber\\
  R_2 &=& \left[i\partial_\mu (\bar d_L \gamma_\mu h_v^+)\right]
		(\bar d_L h_v^-) \nonumber\\
  R_3 &=& \left[i\partial_\mu (\bar d_L h_v^+)\right]
		(\bar d_L \gamma^\mu h_v^-) \nonumber\\
  R_4 &=& \left[iv\cdot \partial (\bar d_L \gamma_\mu h_v^+)\right]
		(\bar d_L \gamma^\mu h_v^-) \nonumber\\
  R_5 &=& i \epsilon_{\lambda\mu\nu\rho} v^\lambda
	  \left[i\partial^\mu	(\bar d_L \gamma^\nu h_v^+)\right]
		(\bar d_L \gamma^\rho h_v^-) \nonumber
\end{eqnarray}
Operators containing $\gamma_5$ or $\sigma$ matrices can be eliminated by the
projection operators implicit in the $d_L$ and $h_v^\pm$ fields.

These operators can mix with each other through the diagrams of fig.2.
In addition, there are also the diagrams of fig.\ 3
which introduce
mixing of the nonlocal operators ${\cal O}_i^+$ with the local operators.
Note that there is no mixing in the other direction since
the local operators do not require the nonlocal ones
as counterterms.

One can employ the symmetries of the effective Hamiltonian to simplify
the mixing matrix. Firstly, the lowest order operator as well as the
$1/m_b$ corrections in (\ref{HQET-heff}) have
the property of being symmetric under Fierz
transformations, i.~e.\ after exchanging the two light quarks
one recovers the original operator by
rearranging the indices of the gamma matrices\footnote{
  For the $R_i$ operators
  which arise in the mixing, in order to obtain the Fierz transform
  a partial integration is also necessary.}.
Therefore it is
useful to switch to an operator basis $\{X_i, Y_j\}$ of definite Fierz parity:
\begin{equation}
  {\cal F}X_i = X_i \quad\quad{\rm and}\quad\quad {\cal F}Y_j = -Y_j.
\end{equation}
In the case at hand, there are seven linear combinations of $P$, $Q$, and $R$
with positive and eight combinations with negative Fierz parity.

Secondly, the operators can be divided into a class of operators
$\{X_i^+, Y_j^+\}$ which are symmetric
under the exchange $v\to -v$, and another class of antisymmetric
operators $\{X^-_i, Y^-_j\}$. It turns out that there is no mixing
between these two sets. Since the
operators introduced by tree-level matching
(\ref{HQET-heff})
have positive parity both under Fierz and $v\to -v$ transformations,
the $X^-$ and $Y^-$ operators can be neglected altogether.

Thirdly, one can eliminate all operators which vanish because of the
equations of motion
\begin{equation}\label{eom}
  v\cdot D h_v^\pm = 0, \quad\quad \slash D d_L = 0
\end{equation}
since they do not contribute to physical matrix elements and do not
mix with operators that do contribute.

The anomalous dimensions of the nonlocal operators is obtained simply
by adding the anomalous dimension of the lowest order effective
Hamiltonian ${\cal O}'_0$ and the anomalous dimensions of the $1/m$ terms
in the effective Lagrangian \cite{FGL91}. Since the former is just
twice the anomalous dimension of the involved current, this pattern
of factorization is found again for the nonlocal part of the recoil
corrections to the four fermion matrix element.

Evaluating the one loop diagrams depicted in fig. 2
and 3  we obtain the result summarized in
tab.~(tab.\ \ref{matrix-table}): The
mixing matrix of the relevant operators
separates into two blocks. The
chromomagnetic moment operator ${\cal O}_3^+$ mixes with the Fierz symmetric
operators
\begin{eqnarray}
  X_1^+ &=& P_2 + P_3 - P_5 + Q_2 + Q_3 - Q_5 \nonumber\\
  X_2^+ &=& 4 R_1 + R_4
\end{eqnarray}
the first one being the operator that already occurred in the matching
conditions (\ref{match}).
The kinetic energy operator ${\cal O}_2^+$ requires as counterterm
a Fierz antisymmetric operator which is renormalized multiplicatively
\begin{eqnarray}
  Y_1^+ &=& 2 R_4.
\end{eqnarray}
With these definitions, the effective Hamiltonian at scales $\mu<m_b$
is given including the $1/m_b$ terms by
\begin{eqnarray}\label{heff-mu}
  H_{\rm eff}
  &=&
  \frac{G_F^2}{\pi^2} |V_{tb}^* V_{td}|^2 m_t^2
          \Phi\left(\frac{m_t^2}{M_W^2}\right)
  \left[ \vphantom{ \frac{1}{m_b}}
  \eta_{\rm QCD}(\mu)\, {\cal O}'_0(\mu)   +\; \frac{1}{m_b}\left\{
   c_2(\mu)\,{\cal O}_2^+(\mu)
 \right.\right.
  \\
  &&\left.\left.+\; c_3(\mu)\,{\cal O}_3^+(\mu)
   + a_1(\mu)\, X_1^+(\mu) +  a_2(\mu)\, X_2^+(\mu)
    + b_1(\mu)\, Y_1^+(\mu)
   \right\}\right].\nonumber
\end{eqnarray}
The array of Wilson coefficients $\{c_k,a_i,b_j\}$ is the solution
of the renormalization group equation
\begin{equation}\label{RGE}
  \mu \frac{d}{d\mu}
  \left(\begin{array}{c} c\\a\\b \end{array} \right)
  + \frac{\alpha_s(\mu)}{12\pi}\,\gamma^{\rm T}
  \left(\begin{array}{c} c\\a\\b \end{array} \right)
  = 0
\end{equation}
where $\gamma$ is the anomalous dimension matrix as listed in
tab.~\ref{matrix-table}.
\begin{table}
  \begin{center}
  $\begin{array}{|l||cc|ccc|cc|}
	\hline
	&{\cal O}_2^+ & {\cal O}_3^+ & X_1^+ & X_2^+ & Y_1^+ \\
	\hline\hline
	{\cal O}_2^+ & 24 & 0 &  0 &   0 & -16 \\
	{\cal O}_3^+ &  0 & 6 &  5 & -30 &   0 \\
	\hline
	X_1^+        &  0 & 0 &  9 &  10 &   0 \\
	X_2^+        &  0 & 0 & -2 &  36 &   0 \\
	Y_1^+        &  0 & 0 &  0 &   0 &  40 \\
	\hline
  \end{array}$
  \end{center}
\caption{The anomalous dimension matrix of the relevant operators. A factor
	 $\alpha_s/12\pi$ has been omitted in all elements.}
\label{matrix-table}
\end{table}

The entries in the anomalous dimension matrix have been calculated
in a general covariant gauge and have been found to be gauge independent,
as it should be the case. However, there is a gauge dependence in the
counterterms of the ${\cal O}_i$ operators corresponding to operators which
vanish according to the equations of motion (\ref{eom}),
indicating that these terms are unphysical.

The solution of (\ref{RGE}) can be written in the form
\begin{equation}\label{formsol}
  \left(\begin{array}{c} c(\mu)\\a(\mu)\\b(\mu) \end{array} \right)
  =
  \exp \left[\frac{1}{2(33-2n_f)} \gamma^{\rm T}
  \ln\frac{\alpha_s(\mu)}{\alpha_s(m_b)}\right]
  \left(\begin{array}{c} c(m_b)\\a(m_b)\\b(m_b) \end{array} \right)
\end{equation}
where $n_f=4$ for $\mu>m_c$, and $n_f=3$ for $\mu<m_c$.
With the initial conditions at
$\mu=m_b$ taken from (\ref{initial}) the solution (\ref{formsol}) reads
\begin{eqnarray}
  c_2(\mu) &=& \zeta(\mu)^{12} \nonumber\\
  c_3(\mu) &=& \zeta(\mu)^{3} \nonumber\\
  a_1(\mu) &=& -\frac{9}{11}\zeta(\mu)^{3}
		-\frac{4(413+\sqrt{649})}{649(-27+\sqrt{649})}
		\zeta(\mu)^{(45-\sqrt{649})/4} \nonumber\\
	   &&	+\;\frac{4(413-\sqrt{649})}{649(27+\sqrt{649})}
		\zeta(\mu)^{(45+\sqrt{649})/4} \\
  a_2(\mu) &=& \frac{14}{11}\zeta(\mu)^{3}
		-\frac{40(295+11\sqrt{649})}{649(27+\sqrt{649})}
		\zeta(\mu)^{(45-\sqrt{649})/4} \nonumber\\
	   &&	+\;\frac{40(295-11\sqrt{649})}{649(-27+\sqrt{649})}
		\zeta(\mu)^{(45+\sqrt{649})/4} \nonumber\\
  b_1(\mu) &=& -\zeta(\mu)^{3}+\zeta(\mu)^{20} \nonumber
\end{eqnarray}
with
\begin{equation}\label{zeta}
  \zeta(\mu) = \left(\frac{\alpha_s(m_c)}{\alpha_s(m_b)}\right)^{1/25}
  \left(\frac{\alpha_s(\mu)}{\alpha_s(m_c)}\right)^{1/27}
\end{equation}
for $\mu<m_c$.
Inserting the values
\begin{equation}
  m_b = 5\;{\rm GeV}, \quad m_c = 1.5 \;{\rm GeV},
  \quad \Lambda_{\rm QCD,4} = 0.2\;{\rm GeV}
\end{equation}
and the one loop approximation for $\alpha_s$, we obtain at $\mu=1\;{\rm GeV}$
\begin{equation}\label{rvector}
  (c_2, c_3, a_1, a_2, b_1)
  = (0.60,\; 0.47,\; 0.52, \;-0.16,\; -0.15)
\end{equation}
together with the value $\eta_{\rm QCD}(1\;{\rm GeV}) = 1.20$.

\section{Estimate of the Matrix Elements}
In order to state numerical results, one has to calculate the
matrix elements of the operators as given above, evaluated between
eigenstates of the lowest-order HQET Langrangian
since the mass dependence resides now in the
coefficients. As long as a non-perturbative calculation
is not available, one has to rely on model assumptions.
We shall estimate all matrix elements using the vacuum insertion
assumption. As pointed out in the introduction we expect this
assumption to at least give the correct order of magnitude since the
parameter $B_B$ calculated from the lattice is indeed of order
unity.

Using the vacuum insertion assumption for
the matrix elements of the effective Hamiltonian we shall
encounter the same matrix elements as in the discussion of the heavy
meson decay constant $f_B$ which has been analyzed to order $1/m_B$
using QCD sum rules \cite{Ne91,Ba91}. The meson decay constant $f_B$ and its
leading recoil corrections may be parameterized in terms of four
constants, which have been estimated in \cite{Ne91}.
Both results for $f_B$  \cite{Ne91,Ba91}
are consistent and we shall use the values given in \cite{Ne91}
for the four parameters.

The parameters describing the heavy meson decay constant
are defined by the following matrix elements
\begin{eqnarray} \label{Fmu}
  \langle 0 | \bar{d} \Gamma h_v^+ | \bbar^0 (v) \rangle
  &=& \frac{1}{2} F(\mu) \,{\rm Tr }\left\{ \Gamma M^+ (v) \right\}\\
  \label{Lbar}
  \langle 0 | i \partial_\mu (\bar{d} \Gamma h_v^+) | \bbar^0 (v) \rangle
  &=& \frac{1}{2} F(\mu) \,\bar{\Lambda} v_\mu\,
  {\rm Tr }\left\{ \Gamma M^+ (v) \right\}
\end{eqnarray}
\begin{eqnarray}
  \label{G1}
  && i \int d^4 x \,
  \langle 0 | \,{\rm T} \left[ (\bar{d} \Gamma h_v^+) (0)\,
            (\bar{h}_v^+ (i D)^2 h_v^+ ) (x) \right]
  | \bbar^0 (v) \rangle \\
  && =  F(\mu) \,G_1 (\mu)\,
  {\rm Tr }\left\{ \Gamma M^+ (v) \right\}\nonumber\\
  \label{G2}
  && i \int d^4 x \,
  \langle 0 | \,{\rm T} \left[ (\bar{d} \Gamma h_v^+) (0)\,
  \frac{g}{2}(\bar{h}_v^+ \sigma_{\mu \nu} G^{\mu \nu} h_v^+ ) (x) \right]
  | \bbar^0 (v) \rangle \\
  &&=
  6 F(\mu) \,G_2 (\mu)\,
  {\rm Tr }\left\{ \Gamma M^+ (v) \right\} \nonumber
\end{eqnarray}
The constant $\bar{\Lambda}$ is the difference between the heavy
quark and the heavy meson mass; the other three constants $F(\mu),
G_1(\mu)$ and $G_2 (\mu)$ depend on the renormalization point $\mu$.
In \cite{Ne91} these constants have been estimated at the low scale
$\mu \sim 2 \bar{\Lambda}$. The values obtained from QCD sum rules
for the four parameters are \cite{Ne91}
\begin{eqnarray} \label{numbers}
  \bar{\Lambda} = 500 \mbox{ MeV}, & \quad \quad &
  F(2 \bar{\Lambda}) = 0.36 \mbox{ GeV}^{3/2} \\
  G_1 (2 \bar{\Lambda}) = -0.5  \mbox{ GeV}, & \quad\quad &
  G_2 (2 \bar{\Lambda}) = - 55 \mbox{ MeV} \nonumber
\end{eqnarray}
where the effective value for $G_1$ as discussed in \cite{Ne91} has
been used.
Furthermore, the matrix $M^+ (v)$ appearing in
eqs.(\ref{Fmu}-\ref{G2}) is the representation matrix for a heavy
pseudoscalar meson containing a heavy quark
\begin{equation}
M^+ (v) = \frac{(-i)}{2} \sqrt{m_B} (1+\slash{v}) \gamma_5.
\end{equation}

Vacuum insertion for the lowest order operator ${\cal O}'_0$
corresponds to the replacement
\begin{equation}
\langle B^0 | {\cal O}'_0 | \bbar^0 \rangle
\to
\frac43\langle B^0 | \bar{d}_L \gamma_\mu h_v^- | 0 \rangle
\langle 0 | \bar{d}_L \gamma^\mu h_v^+ | \bbar^0 \rangle,
\end{equation}
where the factor $4/3$ is a colour factor: The light quark operators
may be contracted in two ways with the light quarks in the mesons
which corresponds to taking the Fierz transform.
One of the possibilities is colour singlet and contributes with
a factor 1, the other one is a combination of colour singlet and octet,
of which only the singlet contributes. This yields in total
a factor of $4/3$ for a Fierz symmetric operator like ${\cal O}'_0$.
Similarly, one would
obtain a factor of $2/3$ for a Fierz antisymmetric operator.

To evaluate this the matrix elements involving mesons containing
a heavy anti-quark are needed. They are given by the same constants
$\bar{\Lambda}, F(\mu),G_1(\mu)$ and $G_2(\mu)$ due to charge
conjugation and are evaluated by replacing the matrix $M^+ (v)$ by
\begin{equation}
M^- (v) = \frac{(-i)}{2} \sqrt{m_B} (1-\slash{v}) \gamma_5.
\end{equation}
Thus we obtain for the matrix element of ${\cal O}'_0$
\begin{equation}
\langle B^0 | {\cal O}'_0 | \bbar^0 \rangle
= \frac13 F^2 (\mu)\, m_B.
\end{equation}

After vacuum insertion all the matrix elements of the local operators
in the order $1/m_B$ $P_i, Q_i$ and $R_i$ may be expressed in terms
of $F(\mu)$ and $\bar{\Lambda}$.
The non-vanishing matrix elements of the local operators are
\begin{eqnarray}
  && \langle B^0 | P_2 | \bbar^0 \rangle
  =  \langle B^0 | Q_3 | \bbar^0 \rangle
  = - \frac{5}{24} \bar{\Lambda} m_B F^2 (\mu)  \nonumber\\
  && \langle B^0 | P_3 | \bbar^0 \rangle
  =  \langle B^0 | Q_2 | \bbar^0 \rangle
  = - \frac{1}{24} \bar{\Lambda} m_B F^2 (\mu)  \nonumber\\
  && \langle B^0 | P_5 | \bbar^0 \rangle
  =  \langle B^0 | Q_5 | \bbar^0 \rangle
  =  \frac{1}{12} \bar{\Lambda} m_B F^2 (\mu)  \\
  && \langle B^0 | R_1 | \bbar^0 \rangle
  = - \frac{7}{24} \bar{\Lambda} m_B F^2 (\mu)  \nonumber \\
  && \langle B^0 | R_2 | \bbar^0 \rangle
  = - \langle B^0 | R_3 | \bbar^0 \rangle
  = - \frac{5}{24} \bar{\Lambda} m_B F^2 (\mu)  \nonumber \\
  && \langle B^0 | R_4 | \bbar^0 \rangle
  =  \frac{1}{6} \bar{\Lambda} m_B F^2 (\mu)  \nonumber
\end{eqnarray}
where the two possible contractions of the light quark in the
four fermion operators have been taken into account. For the local
operators with a definite Fierz parity,
the result is
\begin{eqnarray}\label{local-result}
  \langle B^0 | X_1^+ | \bbar^0 \rangle
  &=& - \frac{2}{3} \bar{\Lambda} m_B F^2 (\mu)  \nonumber\\
  \langle B^0 | X_2^+ | \bbar^0 \rangle
  &=& -  \bar{\Lambda} m_B F^2 (\mu)  \\
  \langle B^0 | Y_1^+ | \bbar^0 \rangle
  &=&   \frac{1}{3} \bar{\Lambda} m_B F^2 (\mu)  \nonumber
\end{eqnarray}

We shall now define the vacuum insertion for the non-local operators
${\cal O}_i^+$. For the piece of ${\cal O}_2$
containing the quark operators $h_v^+$
we define vacuum insertion by
\begin{eqnarray}
  && i \int d^4 x \,
  \langle B^0 | \,{\rm T} \left[ {\cal O}_0 (0)\,
            (\bar{h}_v^+ (i D)^2 h_v^+ ) (x) \right]
  | \bbar^0 (v) \rangle   \\
  &\to&
  \frac43\langle B^0 | \bar{d}_L \gamma_\mu h_v^- | 0 \rangle\,
  i \int d^4 x \,
  \langle 0 | \,{\rm T} \left[ (\bar{d}_L \gamma^\mu h_v^+) (0)\,
            (\bar{h}_v^+ (i D)^2 h_v^+ ) (x) \right]
  | \bbar^0 (v) \rangle \nonumber
\end{eqnarray}
where the colour factor of $4/3$ is the same as in the lowest order piece
${\cal O}'_0$. Thus this matrix element may be expressed in terms of $F(\mu)$
and $G_1 (\mu)$. Analogously we define vacuum insertion for the
matrix element of the non-local operator ${\cal O}_3^+$
and obtain
\begin{eqnarray}\label{nonlocal-result}
  \langle B^0 | {\cal O}_2^+ | \bbar^0 \rangle
  &=&  \frac43 m_B F^2 (\mu)\, G_1 (\mu) \\
  \langle B^0 | {\cal O}_3^+ | \bbar^0 \rangle
  &=&  8 m_B F^2 (\mu)\, G_2 (\mu) \nonumber
\end{eqnarray}
while the matrix elements of ${\cal O}_1$ vanish due to the
equations of motion of the heavy quark.

Within the framework of the vacuum insertion assumption, we now can insert the
expressions for the matrix elements (\ref{local-result}) and
(\ref{nonlocal-result}) into the effective Hamiltonian (\ref{heff-mu}):
\begin{eqnarray}\label{delta-m}
  \Delta M\nonumber
  &=&  \frac{G_F^2}{6\pi^2} |V_{tb}^* V_{td}|^2 m_t^2
          \Phi\left(\frac{m_t^2}{M_W^2}\right) \\
  && \nonumber\times F^2(\mu)
  \left[ \eta_{\rm QCD}(\mu)
  + \frac{\bar\Lambda}{m_b} \left\{4 c_2(\mu)\frac{G_1(\mu)}{\bar\Lambda}
  + 24 c_3(\mu)\frac{G_2(\mu)}{\bar\Lambda}
  \right. \right.\nonumber\\
  &&\quad\left.\left.-\; \vphantom{\frac{\bar\Lambda}{m_b}}
  2 a_1(\mu) - 3 a_2(\mu)  + b_1(\mu) \right\}
  \right].
\end{eqnarray}
With the numerical values taken from (\ref{rvector}) and (\ref{numbers}),
appropriate for the scale $\mu = 1\;{\rm GeV}$, we obtain
\begin{eqnarray}
  \Delta M
  &=& \frac{G_F^2}{6\pi^2} |V_{tb}^* V_{td}|^2 m_t^2
            \Phi\left(\frac{m_t^2}{M_W^2}\right)
  F^2(1\;{\rm GeV})
  \left( 1.20 - 4.4\frac{\bar\Lambda}{m_b} \right) \nonumber\\
   &=& \frac{G_F^2}{6\pi^2 m_B} |V_{tb}^* V_{td}|^2 m_t^2
            \Phi\left(\frac{m_t^2}{M_W^2}\right)
       0.51 \;{\rm GeV}^4
\end{eqnarray}
Thus, with the parameters taken from QCD sum rule analysis,
the $1/m_b$ terms amount to a large correction of about -40\% in this
case. The most important $1/m_b$ corrections arise
from the nonlocal operators; furthermore, the fact that
$G_1$ is not well known \cite{Ne91}
introduces a large uncertainty in the final result.

However, the matrix elements of the nonlocal operators
factorize in leading log order in the same way as the leading
operator ${\cal O}'_0$ does, because the diagrams of
fig. 3
which could lead to non-factorizable contributions only introduce local
operators as counterterms, and the local operators in turn do not
mix with the nonlocal ones. Therefore one can easily absorb these
contributions into the pseudoscalar decay constant $f_B$, if the
contributions of subleading local operators to $f_B$
are taken into account properly.

Explicitly, the square of the pseudoscalar decay constant is given
in leading log
approximation up to order $1/m_b$ by \cite{Ne91,FG90}
\begin{equation}\label{f_B^2}
  f_B^2 = \frac{F^2(\mu)}{m_B} \zeta^{12}(\mu)
  \left( 1  + 2\frac{G_1(\mu)}{m_b}
  + 12 \,\zeta^{-9}(\mu) \frac{G_2(\mu)}{m_b}
  -\frac{\bar\Lambda}{m_b} \left[1 + d(\mu)\right]\right),
\end{equation}
where for $\mu<m_c$
\begin{equation}
  \zeta(\mu) = \left(\frac{\alpha_s(m_c)}{\alpha_s(m_b)}\right)^{1/25}
  \left(\frac{\alpha_s(\mu)}{\alpha_s(m_c)}\right)^{1/27}.
\end{equation}
The radiative corrections in the local term
\begin{equation}
  d(\mu) = \frac{16}{9}(\zeta^{-9}(\mu) - 1 - \ln\zeta^{-9}(\mu))
\end{equation}
remain very small for reasonable values of $\mu$. In particular,
$d(1\;{\rm GeV}) = 0.05$.

Comparing the coefficients of the lowest order operator and the
nonlocal $1/m_b$ terms in (\ref{delta-m}) and (\ref{f_B^2}) shows that
these terms factorize and we have
\begin{eqnarray}
  \eta_{\rm QCD}(\mu)
  \;=\; c_2(\mu) &=& \zeta^{12}(\mu)\, \eta_{\rm QCD}(m_b) \nonumber\\
  c_3(\mu) &=& \zeta^3(\mu)\, \eta_{\rm QCD}(m_b).
\end{eqnarray}
This may be used to rewrite
(\ref{delta-m}) by absorbing the large nonlocal $1/m_b$ contributions
into $f_B^2$:
\begin{eqnarray}
  \Delta M\nonumber
  &=&  \frac{G_F^2}{6\pi^2} |V_{tb}^* V_{td}|^2 m_t^2
          \Phi\left(\frac{m_t^2}{M_W^2}\right)\eta_{\rm QCD}(m_b)\,f_B^2 m_B\\
  && \times
  \left( 1 + \frac{\bar\Lambda}{m_b}
  \left[ 1 + d(\mu) \vphantom{\frac12}\right.\right.
  \\\nonumber
  &&\quad\quad\left.\left.
  + \frac{\zeta^{-12}(\mu)}{\eta_{\rm QCD}(m_b)}(
  -2 a_1(\mu) - 3 a_2(\mu) - a_3(\mu)
  -  b_1(\mu) + b_2(\mu))\right]\right).
\end{eqnarray}
This expression has the form of eq.~(\ref{delta-m-1}) from which we may
read off the bag factor
\begin{equation}\label{bag}
  B_B(m_b) = 1 + 0.45\frac{\bar\Lambda}{m_b} = 1.05
\end{equation}
where $\mu= 1\;{\rm GeV}$ has been taken as the scale
where vacuum insertion is assumed to be valid
for the subleading matrix elements.

In fact, the leading log result only implies that in the static limit
$B_B(\mu)$ is scale independent for $\Lambda < \mu < m_b$, where
$\Lambda$ is the scale where perturbation theory breaks down.
However, the static value $B_B^{\rm stat}$ of $B_B$ is not fixed and we have
simply set
its value to unity. For an arbitrary value of $B_B^{\rm stat}$
our results for $\Delta M$ would be multiplied by this factor.
In particular, eq.\ (\ref{bag}) also is multiplied by $B_B^{\rm stat}$
and thus we have calculated the $1/m_b$ contributions to
$B_B$ as defined in (\ref{delta-m-1}). We note that these corrections
can be calculated perturbatively.

\section{Discussion of the results}
We presented a complete calculation of the leading log QCD corrections
to order $1/m_b$ for the effective Hamiltonian relevant for $B\bbar$ mixing.
The matrix elements have been estimated using the vacuum insertion
assumption which reduces matrix elements of four fermion operators to a
product of current matrix elements between the meson and the vacuum. The latter
have recently been estimated using the QCD sum rule approach \cite{Ne91}.
As far as $f_B$ is concerned the $1/m_b$ corrections according
to \cite{Ne91} are indeed large; however, in our analysis we have
absorbed these large corrections into $f_B$. The remaining corrections
are calculable subleading corrections to the bag parameter which are
of the order $(\alpha_s/\pi) (\bar\Lambda/m_b) \ln(m_b/\mu)$ with
$\mu= 2\bar\Lambda$. They arise only due to QCD effects and hence are small.
We find an enhancement factor for $B_B$ of $1.05$.

We may also compare the results to experiment \cite{Dan91}.
Using a value for $x$
\begin{equation}
  x = \Delta M \cdot \tau_b = 0.67\pm 0.10
\end{equation}
we may study the possible values of $|V_{td}|$ as a function of the top quark
mass. Keeping in mind the above assumptions and using a value of
$f_B$ obtained from the lattice \cite{Lattice}
\begin{equation}
  f_B = 205\;{\rm MeV}
\end{equation}
we find
\begin{equation}
  0.008 \leq |V_{td}| \leq 0.015.
\end{equation}
The range given is due to a variation of the top quark mass between $110$ and
$170\;{\rm GeV}$; however, we have not taken into account the uncertainties
in the $f_B$ determination from the lattice ($\pm 40\;{\rm MeV}$)
and the experimental error in $x$.

Finally, we want to stress that the major part of our calculation
as given in section 2 does not rely on vacuum insertion assumption.
Eventually, the matrix elements of the four fermion operators should be
evaluated non-perturbatively, e.~g.\ by lattice calculations. Since
lattice calculations work best at low scales, our formulae are needed
to connect the low scales with the $b$ mass scale where the matching
to the high energy theory is done.

\section*{Acknowledgements}

We want to thank A.~Buras and P.~Ball for interesting discussions.
We also acknowledge the hospitality of the Physik Department of TU M\"unchen
where this work was completed.

\end{document}